\documentclass[9pt]{article}  \usepackage{times}
\usepackage{graphicx, textgreek}

\usepackage{color}

\usepackage[round,numbers,sort&compress]{natbib} 

\usepackage{amsmath}
\usepackage{amssymb}
\usepackage{stackrel}
\usepackage{chemarrow}
\usepackage{graphicx}
\usepackage{textgreek}

\usepackage{booktabs}
\usepackage{dcolumn}
\usepackage{rotating}
\usepackage{multirow}

\renewcommand\appendix{\par
	\setcounter{section}{0}
	\setcounter{subsection}{0}
	\setcounter{figure}{0}
	\setcounter{table}{0}
	\renewcommand\thesection{Appendix \Alph{section}}
	\renewcommand\thefigure{\Alph{section}\arabic{figure}}
	\renewcommand\thetable{\Alph{section}\arabic{table}}
}

\begin{document}

\twocolumn[{\LARGE \textbf{Nonlinear conductance, rectification and mechanosensitive channel formation of lipid membranes\\*[0.2cm]}}
{\large Karis A. Zecchi and Thomas Heimburg$^\ast$\\*[0.1cm]
	{\small Niels Bohr Institute, University of Copenhagen, Blegdamsvej 17, 2100 Copenhagen \O, Denmark}\\*[-0.1cm]
	
	{\normalsize \textbf{ABSTRACT}\hspace{0.5cm} There is mounting evidence that lipid bilayers display conductive properties. However, when interpreting the electrical response of biological membranes to voltage changes, they are commonly considered as inert insulators. However, lipid bilayers under voltage-clamp conditions do not only display current traces with discrete conduction-steps indistinguishable from those attributed to the presence of protein channels. In current-voltage (I-V) plots they may also display outward rectification, i.e., voltage-gating. Surprisingly, this has even been observed in chemically symmetric lipid bilayers. Here, we investigate this phenomenon using a theoretical framework that models the electrostrictive effect of voltage on lipid membranes in the presence of a spontaneous polarization, which can be recognized by a voltage offset in electrical measurements. It can arise from an asymmetry of the membrane, for example from a nonzero spontaneous curvature of the membrane. This curvature can be caused by voltage via the flexoelectric effect, or by hydrostatic pressure differences across the membrane.
	Here, we describe I-V relations for lipid membranes formed at the tip of patch pipettes situated close to an aqueous surface. We measured at different depths relative to air\slash water surface, resulting in different pressure gradients across the membrane. Both linear and nonlinear I-V profiles were observed. Nonlinear conduction consistently takes the form of outward rectified currents. We explain the conductance properties by two mechanisms: One leak current with constant conductance without pores, and a second process that is due to voltage-gated pore opening correlating with the appearance of channel-like conduction steps. In some instances, these nonlinear I-V relations display a voltage regime in which dI\slash dV is negative. This has also been previously observed in the presence of sodium channels. Experiments at different depths reveal channel formation that depends on pressure gradients. Therefore, we find that the channels in the lipid membrane are both voltage-gated and mechanosensitive. We also report measurements on black lipid membranes that also display rectification. In contrast to the patch experiments they are always symmetric and do not display a voltage offset.
		\\*[0.3cm] }}
\noindent\footnotesize{\textbf{Keywords:} permeability, ion channels, rectification, lipid membrane, flexoelectricity, thermodynamics, voltage-gating, mechanosensitivity\\*[0.1cm]}
\noindent\footnotesize {$^{\ast}$corresponding author, theimbu@nbi.ku.dk. }\\
\vspace{0.3cm}
]

\normalsize


\section{Introduction}
\label{introduction}

The permeability of biological membranes is of immense biological importance. The biological membrane separates inside and outside of cells and has to be selectively permeable to ions and substrates in order to establish well defined chemical potential gradients of the components between inside and outside of the cells. Since this is a formidable task, it is believed that nature must control this by an appropriate ``smart'' mechanism, in particular by providing selective ion channels and pumps to cell membranes \cite[]{Hille1992}. The picture is that of an intelligent pump station with many switches, in which the opening and closing of individual pipes is controlled by substrates or system parameters such as transmembrane voltage, mechanical membrane tension or temperature. Channels that respond to these variables are called voltage-gated channels, mechano-sensitive channels or heat- and cold receptors. A complete field has dedicated its research to the investigation of the molecular nature of the switches. Since there are many substrates and ions, the whole machinery of the biological membrane is complex. In order to understand the working of a membrane with such a complex composition of ``intelligent'' components, one imagines a network of sequential (mostly binary) molecular interactions called pathways. This picture is inherently non-thermodynamic. Instead of making use of thermodynamic variables that act on a complete system with energy, entropy and distributions of states, the channels and receptors seemingly act as receptors to voltage and other intensive variables on the level of single molecules. This picture that does not account for the thermodynamic nature of complex biological ensembles, which must undoubtably exist.

It comes as a profound surprise that appearance of channel-like conduction events can also be seen in pure lipid membranes in the absence of any proteins and macromolecules, i.e. in the complete absence of any single molecule that could act as a channel (e.g., \cite{Antonov1980, Kaufmann1983b, Antonov2005, Blicher2009, Wunderlich2009, Heimburg2010, Blicher2013, Mosgaard2013b}). These channel events are indistinguishable from those of proteins in so far as the current-traces alone do not provide any indication of whether the events originate from proteins or pure lipid membranes. Their single channel conductance, open-lifetime distributions and voltage dependence are very similar to those reported for proteins \cite[]{Blicher2013}. We have called the channels found in lipid membranes ``lipid ion channels'' \cite[]{Heimburg2010} in order to stress these similarities. Lipid channels are thought being due to pores in the lipid membrane that open and close as a consequence of thermodynamic fluctuations. Due to the fluctuation-dissipation theorem, fluctuations become large close to melting transitions. Therefore, in this temperature regime lipid channels and lipid membrane conductance are strongly temperature sensitive \cite[]{Papahadjopoulos1973, Nagle1978b, Sabra1996, Blicher2009}. The lipid membrane permeability is voltage-gated and can display rectified behavior, i.e., the conductance can be largely different at positive and negative voltage \cite[]{Blicher2013, Mosgaard2013b, Mosgaard2015a, Mosgaard2015b}, especially when measured on patch pipettes. Since pure lipid membranes do not contain single macromolecules that could account for the formation of pores, these current fluctuations must be controlled by the thermodynamics of the membrane as a whole. This is striking because one can define a self-consistent macroscopic thermodynamic theory that describes these channels \cite[]{Heimburg2010} without any need for macromolecules.

Such findings represents a serious problem for the interpretation of electrophysiological data. It is easy to demonstrate that quantized conduction events exist in lipid membranes in the absence of proteins. However, it is practically impossible to investigate channel proteins in the absence of membranes. A common approach in electrophysiology is to consider the lipid membrane as an insulator with very high resistance and attribute all discrete opening- and closing events to channel proteins. This is obviously not permissible if the membrane itself can display channels with similar appearance. Many publications have shown that the lipid membrane is not generally an insulator (e.g.,\cite{Papahadjopoulos1973, Nagle1978b, Sabra1996, Blicher2009}). A complete field exists that describes the formation of nano pores in lipid membranes by voltage pulses (electroporation), e.g. \cite{Neumann1999, Boeckmann2008}. This has found clinical applications in drug-delivery and the treatment of cancer \cite[]{Gehl2003, Hoejholt2019}. Thus, the interpretation of current traces in electrophysiological experiments rests on assumptions that are provably not generally true.

While the body of research on protein channels is huge, the permeability of pure membranes is still under-investigated. However, it seems unlikely that on one hand lipid pores represent the sole possible permeation mechanism in synthetic membranes but that on the other hand this mechanism is completely absent in cell membranes. It is interesting to ask the question whether including the thermodynamics of lipid channels into a picture for the biological membrane will help to elucidate the function of cell membranes. Most importantly, it is not known to which degree lipid pores and protein channels may share a similar mechanism or may even be the same. In \cite[]{Mosgaard2013b} we have proposed that proteins could act as catalysts for lipid pore formation, a view that is in line with the experimental finding that truncated proteins that cannot span through the membrane nevertheless can induce pores in biomembranes \cite[]{Stoddart2014}.

Besides the fact that lipid pores display a similar appearance as protein channels, it is not known how they actually look like. \cite{Glaser1988} have proposed that there exist hydrophobic and hydrophilic pores with openings on nanoscale, a view that is consistent with molecular dynamics simulations \cite[]{Boeckmann2008}.

In this paper we study I-V profiles of synthetic lipid membrane patches using patch-clamp recordings. We use the droplet-technique, in which the membrane is formed across a patch pipette that is in contact with the aqueous surface of a buffer \cite[]{Hanke1984, Gutsmann2015}. In this setup, the depth of the pipette can be altered. This will influence the pressure gradients across the membrane. We study the depth dependence, analyze the theory of the I-V profiles and combine it with theoretical consideration about the equilibration processes directly after a voltage jump. We compared the patch experiments with black lipid membrane (BLM) experiments which are performed on much larger membranes. Finally, we compare our findings with potassium channels.

\section{Materials and methods}
\label{materialsandmethods}

\subsection{Lipids and Chemicals}
\label{lipidsandchemicals}

1,2-dimyristoyl-\emph{sn}-glycero-3-phosphocholine (DMPC), 1,2-dilauroyl-\emph{sn}-glycero-3-phosphocholine (DLPC) and cholesterol were purchased from Avanti Polar Lipids (Alabaster\slash AL, US), stored in a freezer and used without further purification. Lipid patches consisted of DMPC:DLPC=10:1 (mol:mol) for the patch clamp experiments and of DMPC:DLPC:choles\-terol = 77.3:7.7:15 (mol:mol:mol) for the black lipid membrane experiment. Each lipid was suspended separately in chloroform and then mixed to the desired ratio. The mixture was then dried under vacuum. In the patch experiments, the dry lipids were resuspended in Hexane:Ethanol=4:1(mol:mol) to a final concentration of 2mM for the patch clamp experiment. The cholesterol mixture used in the BLM experiments was dissolved in decane to a final concentration of 10mg\slash mL.

In the patch clamp experiment, the electrolyte solution used on both sides of the membrane consisted of 150 mM KCl, 150 mM NaCl and it was buffered with 50 mM TRIS to a final pH of 7.6. All water used in the experiments was purified with a Direct-Q Water Purification System (Merck Millipore, Germany) and had a resistivity larger than 18.1 M$\Omega\cdot$cm. In the BLM experiments we used 150 mM KCl, 150 mM NaCl, 2 mM HEPES and 1 mM EDTA (both from Sigma-Aldrich, Germany), pH was adjusted to 7.4

\begin{figure}[htbp]
	\centering
	\includegraphics[width=250pt]{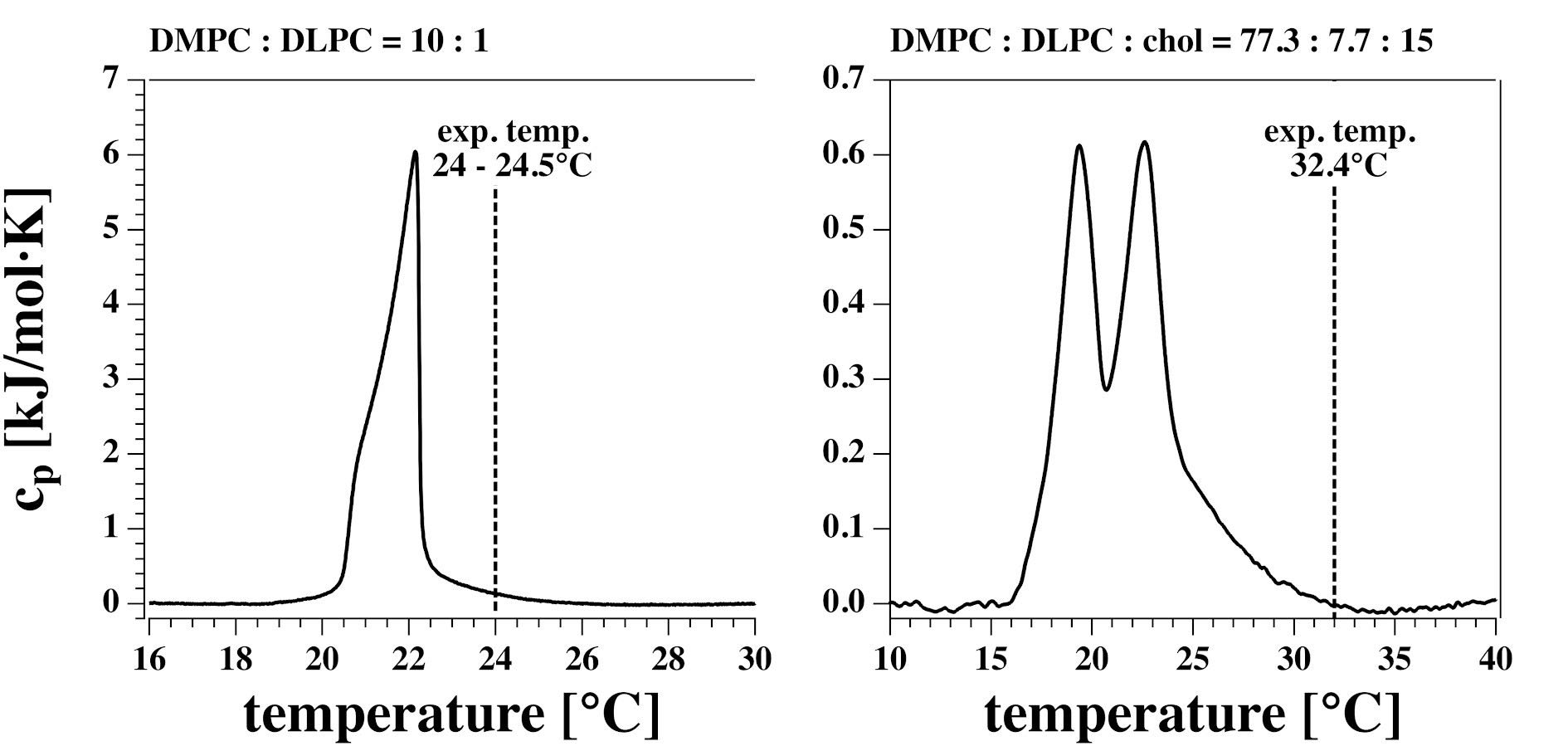}
	\caption{\small\textit {Calorimetric profiles of the two lipid mixtures used. The vertical dashed line indicates the experimental temperature at which I-V profiles and currents were measured.}}
	\label{figure1}
\end{figure}

Experiments were performed at a temperature at the upper end of the melting regime for all lipid mixtures used (see Fig. \ref{figure1}).

\subsection{Methods}
\label{methods}

\textbf{Patch clamp experiments:} Glass micropipettes were pulled from borosilicate capillaries with a vertical PC-10 puller (Narishe Group, Japan) following the two-steps procedure explai\-ned in \cite{Laub2012}. They were then fire-polished using a Narishige MF-900 Microforge, which created pipette openings on the order of 10 \textmu m.

Lipid membrane patches were reconstituted on the tip of glass pipettes following the method introduced by \cite{Hanke1984} and described in detail in \cite{Laub2012, Gutsmann2015}. According to the protocol a droplet of lipid solution is placed on the outer wall of a glass micropipette filled with the electrolyte. The pipette stands vertically with its tip in contact with the liquid\slash air interface of a buffered electrolyte filled glass beaker. As the lipid droplet flows down to the pipette tip, it gets sealed by a spontaneously formed lipid bilayer.

Two Ag\slash AgCl electrodes were placed one inside the pipet\-te and the other one in the bulk electrolyte, the latter acting as ground electrode. They were both connected to a patch clamp amplifier (Axopatch 200B, Molecular Devices, US) through a headstage to which the pipette was also secured. The amplifier was run in whole cell mode, the signal was sampled at a frequency of 10 kHz and filtered with a 2 kHz low pass Bessel filter. The headstage was allowed to move vertically with the aid of a micromanipulator (Luigs $\&$ Neumann, Germany), with which the vertical position of the tip relative to the electrolyte surface could be monitored.

\textbf{Black lipid membrane experiments:} A DMPC:DLPC: cholesterol = 77.3:7.7:15 (mol:mol:mol) mixture was dissol\-ved in decane to a final concentration of 10 mg\slash mL. Black lipid membranes were formed on a circular aperture in a 25 \textmu m thick Teflon film. We used commercially available horizontal bilayer slides (Ionovation GmbH, Germany) made of two microchambers (filled with approx 150 \textmu l of the same electrolyte solution) separated by an horizontal Teflon film. The upper and lower chambers are connected only through the 120 \textmu m aperture in the film. Once a small droplet ($\approx$ 0.2 \textmu l) of lipid solution is placed in the upper chamber close to the aperture, a bilayer is formed automatically by a microfluidic perfusion system (Ionovation Explorer, Ionovation GmbH, Germany). The membrane formation was monitored with capacitance measurements and was automatically repeated until the membrane capacitance was stable above a minimum threshold value of 40 pF. The bilayer slide was placed on the work stage of a in inverted microscope (IX70, Olympus, Japan) which allowed for optical monitoring of the bilayer formation. See \cite{Zecchi2017} for more details.

\textbf{Differential scanning calorimetry:} Heat capacity profiles were obtained using a VP scanning-calorimeter (MicroCal, Northampton, MA) at a scan rate of 5$^\circ$\slash h.

\subsection{Electrophysiological experiments}
\label{electrophysiologicalexperiments}

\begin{figure}[htbp]
	\centering
	\includegraphics[width=224pt,height=143pt]{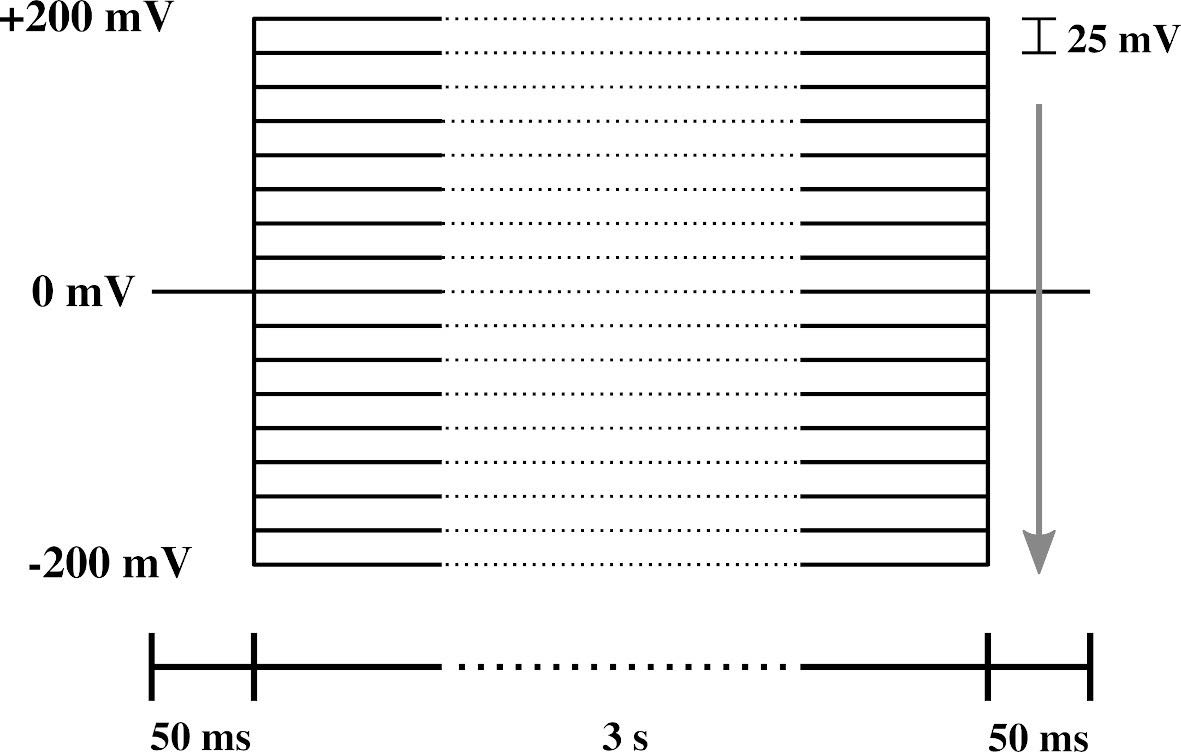}
	\caption{\small\textit {Protocol of the voltage steps used in the I-V measurement. The lower bar indicates the duration of each part of the protocol. The arrow shows the direction of voltage steps, which ranged from $200 \,$ mV to $-200\,$ mV in steps of $-25\,$ mV.}}
	\label{figure2}
\end{figure}

All experiments are representative and qualitatively reproducible. However, membranes break easily. In patch clamp measurements, voltage-jumps were performed between 200 mV and -200 mV in steps of 25 mV (Fig. \,\ref{figure2}). Each step lasts 3.1 seconds for 17 different voltage-clamp traces needed for one I-V profile. Thus, each series lasts about 53 seconds. Only few membranes are stable long enough for a complete series of voltage jumps that lead to a single I-V profiles. Even less membranes allow for recording several I-V profiles to check for reproducibility and the variation of pipette depth. The typical interval between two I-V profiles recorded on the same membrane is 1--5 minutes. The series of I-V profiles shown in Fig. \,\ref{figure6} (top) contains 5 traces for one single depth of the pipette in the aqueous medium, which corresponds to 10--25 minutes. Since we could measure on this membrane at 3 different depths, the membrane was stable for a total of $\sim$ 30 min - 1 hour. For this reason, all patch clamp data shown in this paper originate from two different membrane patches that were sufficiently stable to not break during many I-V recordings. We name them membrane 1 and membrane 2 throughout the text. However, we have many more experiments that are consistent with our results, where the membranes did not last long enough for an extended series of recordings.

The voltage-jump protocol for the BLM measurements shown in Fig. \ref{figure10} was different. Here, we changed the voltage from +10 mV to -10 mV, then to 20 mV and -20 mV and so on (not shown).

\section{Results}
\label{results}

\subsection{Description of the experiment}
\label{descriptionoftheexperiment}

\begin{figure}[htbp]
	\centering
	\includegraphics[width=250pt]{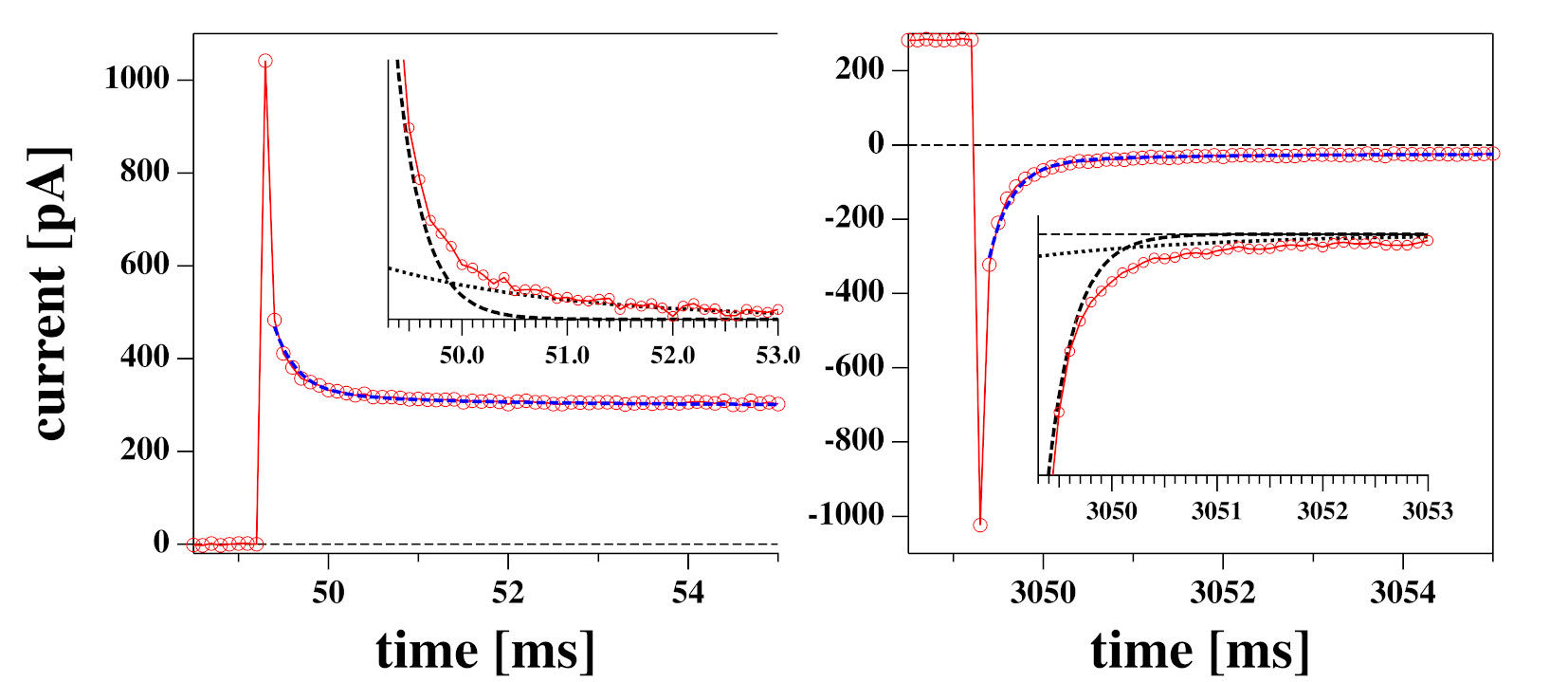}
	\caption{\small\textit {Initial phase of a voltage jump from 0 V to 0.175 V (left) and back (right) from the experiment shown in Fig. \,\ref{figure4} B and D. The profiles are reasonably well fitted by a biexponential function with the same two relaxation times for the jump in both directions. The two exponential functions are shown in the inserts.}}
	\label{figure3}
\end{figure}

After a voltage jump, all the current traces measured showed an initial transient decay of about 10 ms from a current peak at $t=50$ ms (the time of the voltage step, see Fig.\ref{figure3}) to their steady state value (the current value at $t=3$ s). The transient part of the current contains information about the charging of the pipette (and electrodes), the charging of the membrane and any relaxation phenomena in the membrane which can lead to changes in resistance, capacitance and polarization.

\begin{figure*}[htbp]
	\centering
	\includegraphics[width=350pt]{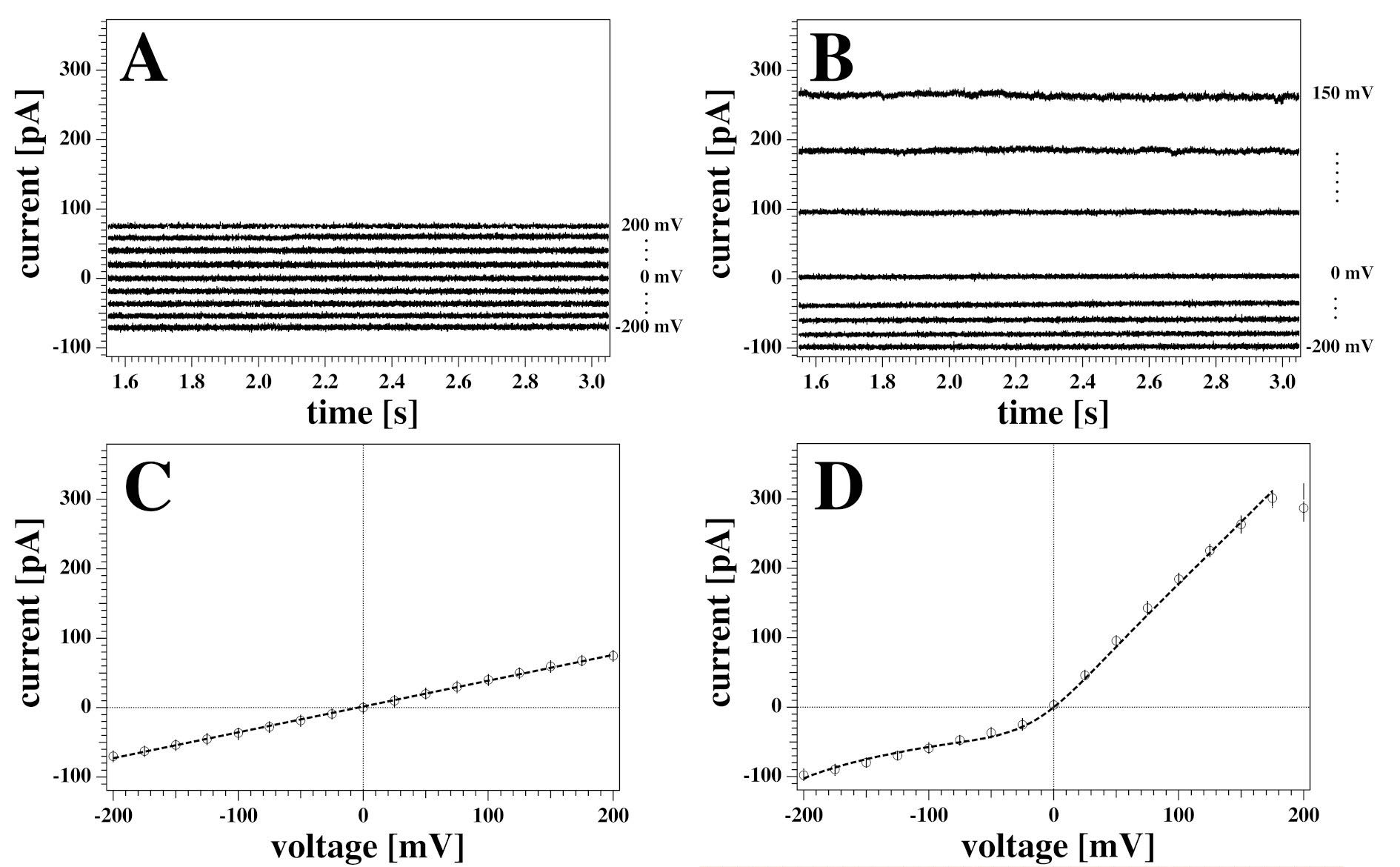}
	\caption{\small\textit {\textbf{A-B:} Detail of the last $1.5\,$ s of two representative current responses to the voltage pulses shown in Fig. \,\ref{figure2} recorded from the same membrane. The graph shows the response to only every other voltage step for clarity. Current traces were corrected for the initial offset at zero voltage. Currents in A were measured with the tip of the pipette at the air\slash water interface. Traces in B were measured with the tip of the pipette $3\,$ mm below the water surface. \textbf{C-D:} Current-voltage relationship for the traces in A-B, respectively. All points of all the current traces between $1.5\,$ s and $3\,$ s were plotted and fitted. The model used for the fit were a linear relation in panel C, and eq.\,\ref{eq:theor1.5} in panel D as outlined in the Theory Section. The fit gives a conductance of $g_L =372 \pm 4$ pS for the recording in $A$ and $C$, and a voltage offset of $V_0=209$ mV for the recording in $B$ and $D$. Solid circles show the average values of the current. The measurements were performed at a temperature of $T=297.85$ K. }}
	\label{figure4}
\end{figure*}

The relaxation processes are well described by a biexponential function (Fig.\ref{figure3}). We decided to not include the first data point in the fit. This corresponds to the first 100 \textmu s which is the time resolution of our experiment.

The current response of the membrane after a voltage jump was measured. As an example, we show two sets of recordings in Fig.\ref{figure4} A, B. Possible voltage offsets were corrected by subtracting the mean value of the current at the holding voltage from the corresponding current trace during the step protocol.

In order to obtain steady state currents for the I-V profiles, we restricted ourselves to determine averages of the second half of the current traces corresponding to the last 1.5 s after each voltage step (open circles in Fig.\ref{figure4} C, D).

\subsection{General trends}
\label{generaltrends}

Both recordings in Fig.\ref{figure4} A and B were obtained from the same membrane (membrane 1). Nevertheless, they show distinct features that are representative of all the recordings made on the two membranes used in the patch clamp measurements described in this work. These are outlined below.

Fig.\ref{figure4} A shows the current response of the membrane when the tip of the pipette was at the air\slash water interface. The current traces are symmetric with respect to voltage and their relatively small value indicates a large membrane resistance. This can be quantified by inspection of the correspondent current-voltage relationship (Fig.\ref{figure4} C). A linear fit to the data in Fig.\ref{figure4} C gives a value for the conductance of $ g_L$=(372$\pm$4) pS, or equivalently, a membrane resistance of about $R_0$=2.69 G$\Omega$. The conductance in this case is constant and independent of voltage.

A different scenario is shown in the traces plotted in Fig.\ref{figure4} B, measured for the same membrane at a different depth with respect to the water surface (3mm). Here, the current response to positive and negative voltage-jumps is clearly different, as confirmed by the asymmetric and nonlinear I-V curve in Fig.\ref{figure4} D. The membrane appears to be more conductive at positive voltages, showing a fairly constant conductance of about $g=1.7$ nS (or resistance of roughly $R=500$ M$\Omega$), as obtained by a linear fit to the positive voltage range. Thus, it is almost five times larger than the linear I-V profile shown in Fig.\ref{figure4} C. The response to negative voltages, in contrast, is slightly nonlinear and less pronounced, comparable in magnitude to the linear case.

The data point at $200$mV in Fig.\ref{figure4} D has not been included in the fit. In our protocol this is the first datapoint. The corresponding current trace shows a transient behaviour and does not equilibrate fully in the 3 seconds of the test pulse. This transient behaviour of the current starting from a low conductance value and increasing without reaching a steady state was not uncommon. It was only observed at positive high voltages, and under few instances (in case of reversed voltage protocol) at high negative voltages. In the absence of any satisfactory explanation, in this work traces showing similar behaviors were discarded from steady-state analysis. Further, we generally find that the first datapoint in each I-V profile is an outlier with respect to an otherwise systematic behavior. This might also be related to an equilibration of the membrane patch after the first voltage jump of a series.

The two sets of recordings shown in Fig.\ref{figure4}A and B have been performed at different depths of the pipette tip with respect to the bath surface. Different vertical positions of the pipette relative to the surface correspond to different values of hydrostatic pressure at the bath-facing leaflet of the membrane. Since the pressure at the inner leaflet is constant, this corresponds to different pressure gradients across the membrane. A pressure difference between the two leaflets can result into a net curvature according to the Young-Laplace law. Curvature in a chemically symmetric membrane can break the otherwise symmetric charge and dipole distribution, and therefore produce a voltage offset, as outlined in \cite{Mosgaard2015a}.

\subsection{Theoretical considerations}
\label{theoreticalconsiderations}

\subsubsection{Current-voltage relations}
\label{current-voltagerelations}

The general tendency of a membrane to display a higher conductance at positive as compared to negative voltages is known as outward rectification. In the case of biological membranes, it's customary to ascribe electrical rectifications like the one shown here to the voltage dependent behaviour of certain protein-channels spanning the membrane. However, outward rectified I-V curves like the one of Fig.\ref{figure4} D have already been observed earlier in protein-free membranes \cite[]{Laub2012, Blicher2013}. In these publications, the rectified behavior was explained on the basis of a capacitor model like the one introduced below. It requires the formation of membrane pores and a spontaneous electrical membrane polarization as caused from an asymmetry between the two monolayers of a bilayer \cite[]{Mosgaard2015b}. This could, for instance, originate from membrane curvature that changes the relative lipid dipole density on the two monolayers of the membrane.

We will describe the conductance of a membrane by using a description from \cite{Mosgaard2015a}.

We assume that the membrane contains pores with an open probability that display a quadratic voltage dependence \cite[]{Winterhalter1987, Blicher2013}. The free energy of an open pore is given by

\begin{equation}
\Delta G=\Delta G_0+\alpha \left[(V+V_0)^2-V_0^2\right]
\label{eq:theor1.1}
\end{equation}

Here, $\Delta G_0$ is the free energy of a pore at a voltage of $V=0$ V, and $\alpha$ is a coefficient. $\Delta G$ displays a minimum at $V=-V_0$, where $V_0$ is the voltage offset that originates from a polarization of the membrane. Its origin will be discussed below. Defining an equilibrium constant $K=\exp(-\Delta G/kT)$ between a closed and an open pore, we obtain for the probability of finding an open or a closed pore, $P_{open}$ or $P_{closed}$, respectively:

\begin{equation}
P_{open}=\frac{K}{1+K} \quad \mbox{and} \quad P_{closed}=\frac{1}{1+K}
\label{eq:theor1.2}
\end{equation}

which sum up to one. If we assume that conduction occurs exclusively via open pores in the membranes, the overall current through the membrane pores will be given by

\begin{equation}
I_p=g_{pore} N P_{open} V\equiv g_p P_{open}V\;,
\label{eq:theor1.3}
\end{equation}

where $g_{pore}$ is the conductance of a single open pore, $N$ is the total number of pores, and $g_p=N g_{pore}$ is the conductance of $N$ open pores.

In the experimental section we find that there exist current traces that do not display any open pore events. This is mostly the case when the current-voltage relation is linear. It is therefore possible that there exists a voltage-independent leak current, $I_L$, and voltage-dependent current through pores in a membrane, $I_p$. If this were the case, the current-voltage relation in eq. \,\ref{eq:theor1.3} would be given by

\begin{equation}
I=I_L+I_p=\left(g_L+g_p P_{open} \right) V
\label{eq:theor1.4}
\end{equation}

where $g_L$ is a constant leak-conductance of the membrane in the absence of pores. This equation assumes identical ion concentrations on both sides of the membrane (Nernst potential $E_0$ is zero). If the ion concentrations were different from zero, we would obtain

\begin{equation}
I=\left(g_L+g_p P_{open} \right) (V-E_0)
\label{eq:theor1.5}
\end{equation}

with $E_0=(RT/zF)\ln(c_{out}/c_{in})$ for an ion with charge $z$ and the concentrations $c_{out}$ and $c_{in}$ outside and inside of the pipette, respectively. Since we use the same buffer in the pipette and in the external medium, the Nernst potential in our experiments is $E_0=0$ V.

\begin{figure*}[htbp]
	\centering
	\includegraphics[width=454pt,height=138pt]{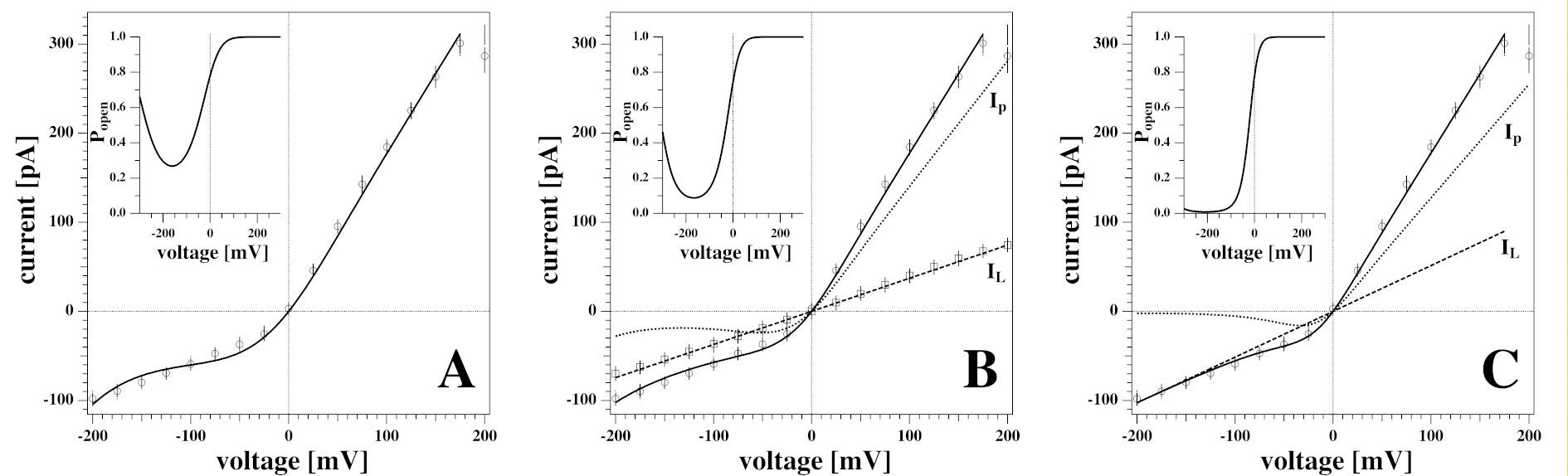}
	\caption{\small\textit {Three different scenarios to fit the nonlinear current-voltage relations. \textbf{A:} conductions by pores only ($g_L=0$). \textbf{B:} Conduction by voltage-dependent pores and a voltage-independent fixed leak current taken from the experiment in Fig. \,\ref{figure4} C. \textbf{C:} Same as B but the leak current was not fixed. The latter fit describes the data best. The inserts show the pore open probability for the three different scenarios. The minimum of $P_{open}$ represents the offset voltage $-V_0$. See text for details.}}
	\label{figure5}
\end{figure*}

Fig. \,\ref{figure5} shows the non-linear rectified I-V profile from Fig. \,\ref{figure4} D and three attempts to describe it. We expect the I-V profile to pass through zero ($E_0=0$) because we have the same ion concentrations on both sides of the membrane. We corrected for slight deviations in the current at zero voltage. For the fit in Fig. \,\ref{figure5} A we assumed that there is only one single permeation process by pores and no leak-conductance (ie, $g_L=0$ in eq. \,\ref{eq:theor1.5}). The fit is reasonable but not perfect. The insert shows the calculated pore open-probability. It displays a minimum at $-162$ mV corresponding to a spontaneous polarization of the membrane that leads to a voltage offset of $V_0=+162$ mV. The minimum open-probability at this voltage is about $0.27$ which requires 27 \% of all pores being open. However, we will see below that no open pores can be detected at this voltage. For this reason we decided that this is not the most likely scenario. Fig. \,\ref{figure5} C shows a free fit allowing for a variation of the leak conductance $g_L$. The straight line in this panel corresponds to the leak currents $I_L=g_L·V$. In this fit, the open probability of the pores reaches a minimum at -208 mV and the minimum open probability of pores is below 1 \%. Since there is one more fit parameter, it is not surprising that this describes the measured I-V profile better. However, what also speaks in favor of this description is that one does not expect open pores at negative voltages. Only at positive voltages, the open probability of pores is significantly different from zero. The fit in Fig. \,\ref{figure5} C is composed of a leak current $I_L$ and a pore current $I_p$, which are independently shown in the figure.

In our experiments we sometimes find linear I-V relations and sometime outward rectified profiles. It is not exactly clear why both cases occur with the same experimental settings for the same membrane. Interestingly, the linear I-V profile (Fig. \,\ref{figure4} C and Fig. \,\ref{figure5} B) yields a quite similar conductance $g_L$ as leak current $I_L$ in Fig. \,\ref{figure5} C. This supports our assumption that the conductance of the lipid membranes is a phenomenon described by two different processes: A voltage-independent leak conductance and a voltage-gated pore formation process.

Therefore, we tentatively assumed that the two cases are only distinguished by the presence or absence of prepores that are ready to open but that the leak conductance of the membrane is identical in all experiments. Fig. \,\ref{figure5} B shows a fit where the leak conductance $g_L$ was obtained from the linear I-V profile shown in Fig. \,\ref{figure4} C. Its value was kept constant for fitting the rectified profile. We see that this describes the I-V profile quite well.

\subsubsection{Relaxation processes}
\label{relaxationprocesses}

In Fig. \,\ref{figure3} one could see that the initial current during equilibration displays more than one exponential component. Time-dependent changes of the membrane current can have two origins. The first is the charging of the membrane capacitor, of pipette walls and electrodes. The charge on a capacitor is given by

\begin{eqnarray}
q &= & C_m V+ A\cdot P_0\;,
\label{eq:theor2.1}
\end{eqnarray}

where $C_m$ is the capacitance, A is the area of the capacitor and $P_0$ is its spontaneous polarization \cite[]{Mosgaard2015a}, which is related to the voltage offset $V_0$ described above. The capacitive current is therefore given by

\begin{equation}
I_c(t) = C_m\frac{dV}{dt} +V\frac{C_m}{dt} +\frac{d}{dt} (A\cdot P_0)
\label{eq:theor2.2}
\end{equation}

The first term on the right hand side is considered in electrophysiology, while the second and third term are neglected. Thus, in the textbooks it is assumed that the capacitance of the membrane and all membrane properties are constant after a voltage jump. This assumption requires that the membrane structure is independent of voltage, which is practically impossible for a soft molecular layer the will deform in the presence of electrostatic forces. The time constant of charging a constant capacitor in an electrolyte solution, $\tau_0$, is typically fast because it is dominated by the low resistance of the electrolytic medium. The second term in eq. \,\ref{eq:theor2.2} represents the time-dependent change in capacitance caused by a voltage-induced structural change in the membrane, and the third term is the related voltage-induced change in the spontaneous polarization of the membrane, for instance caused by changes in lipid head-group orientation or changes in curvature. Due to electrostriction, the capacitance of membranes is voltage-dependent \cite[]{Heimburg2012, Mosgaard2015a, Mosgaard2015b}. The capacitance of a membrane is given by $C_m=\varepsilon_0\varepsilon A/D$ where $A$ is the membrane area and $D$ is the membrane thickness. Electrostriction reduces the membrane thickness and increases the membrane area. Both effects lead to an increase in capacitance, $\Delta C_m$. This effect is most pronounced close to melting transitions in membranes because here the membranes are softest. This is the situation treated in this paper (see Fig. \ref{figure1}). In this paper we assume that the slow timescale of capacitance and polarization changes, $\tau_m$, results from the relaxation processes in membranes, which are in the millisecond regime \cite{Grabitz2002, Seeger2007}.

A further time-dependent change in the membrane current may originate from voltage-induced changes in the membrane conductance, $\Delta g$, due to changes in membrane structure. It is known that membranes are more permeable in their melting transitions \cite[]{Papahadjopoulos1973, Nagle1978b, Sabra1996, Blicher2009}. Therefore, a voltage-induced change in membrane structure as caused by electrostriction will also change the conductance of the membrane. In the presence of a spontaneous voltage-offset (polarization) of the membrane, this effect will be different for positive and negative voltages, i.e., it will be rectified. Since it is related to structural changes in the membrane, the time-scale of its changes will also be dictated by the relaxation time-scale in the membrane, $\tau_m$, where we have assumed a single-exponential relaxation process (as described in \cite{Grabitz2002, Seeger2007}).

For the total membrane current we obtain for a jump from voltage $V_b$ before the jump to a voltage $V_a$ after the jump:

\begin{eqnarray}
I(t) &=&\frac{C_{m,b}}{\tau_0}\Delta V e^{-\frac{t}{\tau_0}} +\left(\Delta C_m V_a + \Delta (A P_0)\right) \frac{e^{-\frac{t}{\tau_m}}}{\tau_m}   \nonumber \\
&&+\left( g_b + \Delta g \left(1-e^{-\frac{t}{\tau_m}}\right) \right)V_a  =\\
&=& g_a V_a + \frac{C_{m,b}}{\tau_0} \Delta V e^{-\frac{t}{\tau_0}} + \nonumber\\
&&\left(\frac{\Delta C_m V_a + \Delta (A P_0)}{\tau_m} -\Delta g V_a \right) +e^{-\frac{t}{\tau_m}}=\nonumber\\
&=&g_a V_a+ A_0 e^{-\frac{t}{\tau_0}} + A_m e^{-\frac{t}{\tau_m}}\;,\nonumber
\label{eq:theor2.3}
\end{eqnarray}

where the term related to charging the capacitor is described by the timescale $\tau_0$ and the amplitude $A_0$. All terms related to changes in membrane structure change with the time constant $\tau_m$ and amplitude $A_m$. $C_{m,b}$, $C_{m,a}$, $g_b$ and $g_a$ are the steady-state capacitance and conductance before and after the voltage jump, respectively. $\Delta V=V_a-V_b$, $\Delta C_m=C_{m,a}-C_{m,b}$, $\Delta g = g_a-g_b$ and $\Delta(A P_0)$=$(A P_0)_a$ - $(A P_0)_b$ are the differences of the respective functions in steady state before and after the voltage change. We see that the steady-state current after a jump is given by

\begin{equation}
I (V_a)= g_a V_a\;.
\label{eq:theor2.4}
\end{equation}

The time-dependent current contributions are dominated by two time-scales. One of them, $\tau_0$, is related to charging a constant capacitor, while the second one, $\tau_m$, is slow and dominated by the relaxation process of conductance, capacitance and polarization of the membrane.

\begin{figure*}[htbp]
	\centering
	\includegraphics[width=336pt,height=409pt]{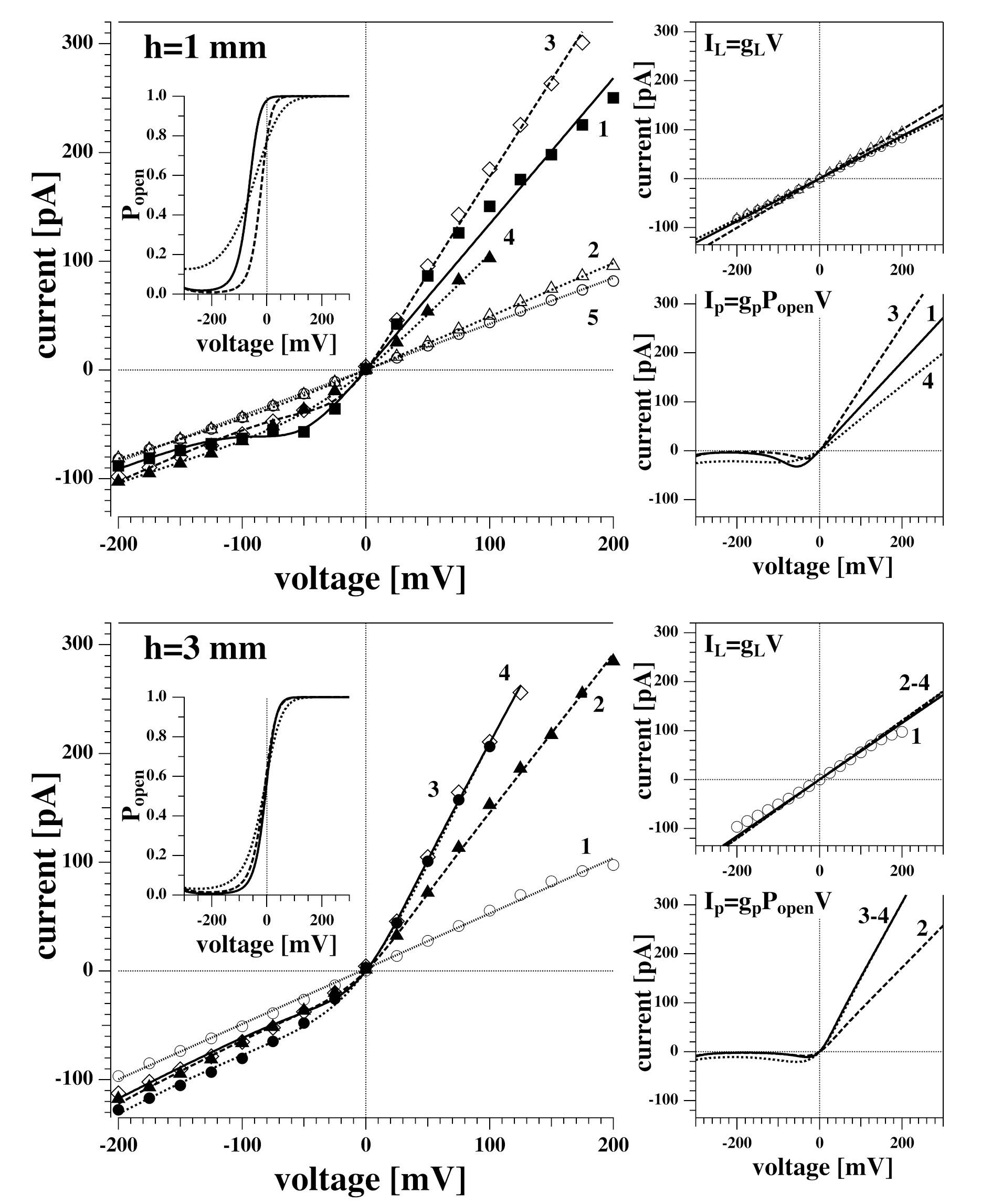}
	\caption{\small\textit {I-V recordings of the same membrane as Fig.\ref{figure4} (membrane 1) at different depths below the air\slash water interface. \textbf{Top:} Tip $1\,$ mm below the surface. \textbf{Bottom:} Tip $3\,$ mm below the surface. Both sets of recordings show two different electrical behaviors, with both linear and nonlinear I-V relationships. The curves were measured in sequence and are numbered in order of appearance. Only the average values of the current are displayed for clarity. The dotted lines are fit to all current data points. }}
	\label{figure6}
\end{figure*}

\subsubsection{Experiments at different pipette depths}
\label{experimentsatdifferentpipettedepths}

In order to better understand the appearance of the nonlinear behaviour and the origin of the voltage offset, I-V measurements were performed on the same membrane at different positions of the tip in the water bath. This is an indirect attempt of controlling the hydrostatic pressure gradient across the membrane, which increases linearly with the depth, $h$.

With the aid of a micromanipulator, the tip of the pipette was lowered from its initial position (close to the water surface) down to different depths inside the water bath. Fig.\,\ref{figure6} shows two series of I-V curves measured for the same membrane (membrane 1) at different positions, i.e., at 1 and 3 mm below the water surface, corresponding to a pressure difference of about 98 and 294 Pa. The numbers close to each curve indicate the temporal order of the recording in each sequence. The time interval between subsequent recording was not fixed but was never more than 5 minutes (with an average of 1 minute and a half).
Fits for the I-V profiles were generated by using eq. \,\ref{eq:theor1.4} and the procedure used in Fig. \,\ref{figure5}C. The two contributions to the conductance are displayed separately in the small panels. No qualitative differences can be observed between the recordings at 1 and 3 mm. Both positions produce consistently both linear and nonlinear responses, the latter being always in the form of outward rectified I-V curves. Interestingly, during each voltage-jump sequence, one behavior was consistently maintained while in the next sequence one can observe a different behavior. For instance, in Fig.\,\ref{figure6} top left, the first trace was outward rectified, the second was linear, the 3rd and 4th trace were rectified and finally the 5th was linear. The reason for this behavior is not clear. It seems that the linear contribution of the conductance is the same in both, linear and rectified profiles - but that it is not always possible to activate pores.  The values of the voltage offset as obtained from the fit vary slightly from one recording to the other, but they are comparable between the two different positions. On average, it is 243 ( $\pm$ 34) mV at 1 mm depth and 221 ( $\pm$ 20) mV on average at a depth of 3 mm. The leak conductance $g_L$ and the pore conductance $g_p$ were larger at larger depth of the patch pipette. For membrane 1, the conductance $g_L$ increased by 32\% and the pore conductance $g_p$ by 37\% when going from 1 to 3 mm depth.

It is interesting to note that trace 1 in the top left panel of Fig. \,\ref{figure6} displays a voltage regime around $-50$ mV, in which the dependence of the current on voltage, $dI/dV$, is negative. This is impossible without a voltage-dependent conductance. It can be explained if one considers that around -200 mV, all pores are closed while at -50 mV some pores are open. For this reason, one can find a larger negative current at -50 mV than at -200 mV.

\subsubsection{Channel activity}
\label{channelactivity}

Nonlinear I-V curves and outward rectification are not the only properties of the lipid membranes studied here that resembled those of biological membranes. We also find that several recordings of membranes showed current fluctuations and quantized steps that are typical of ion channel activity.

\begin{figure*}[htbp]
	\centering
	\includegraphics[width=397pt,height=390pt]{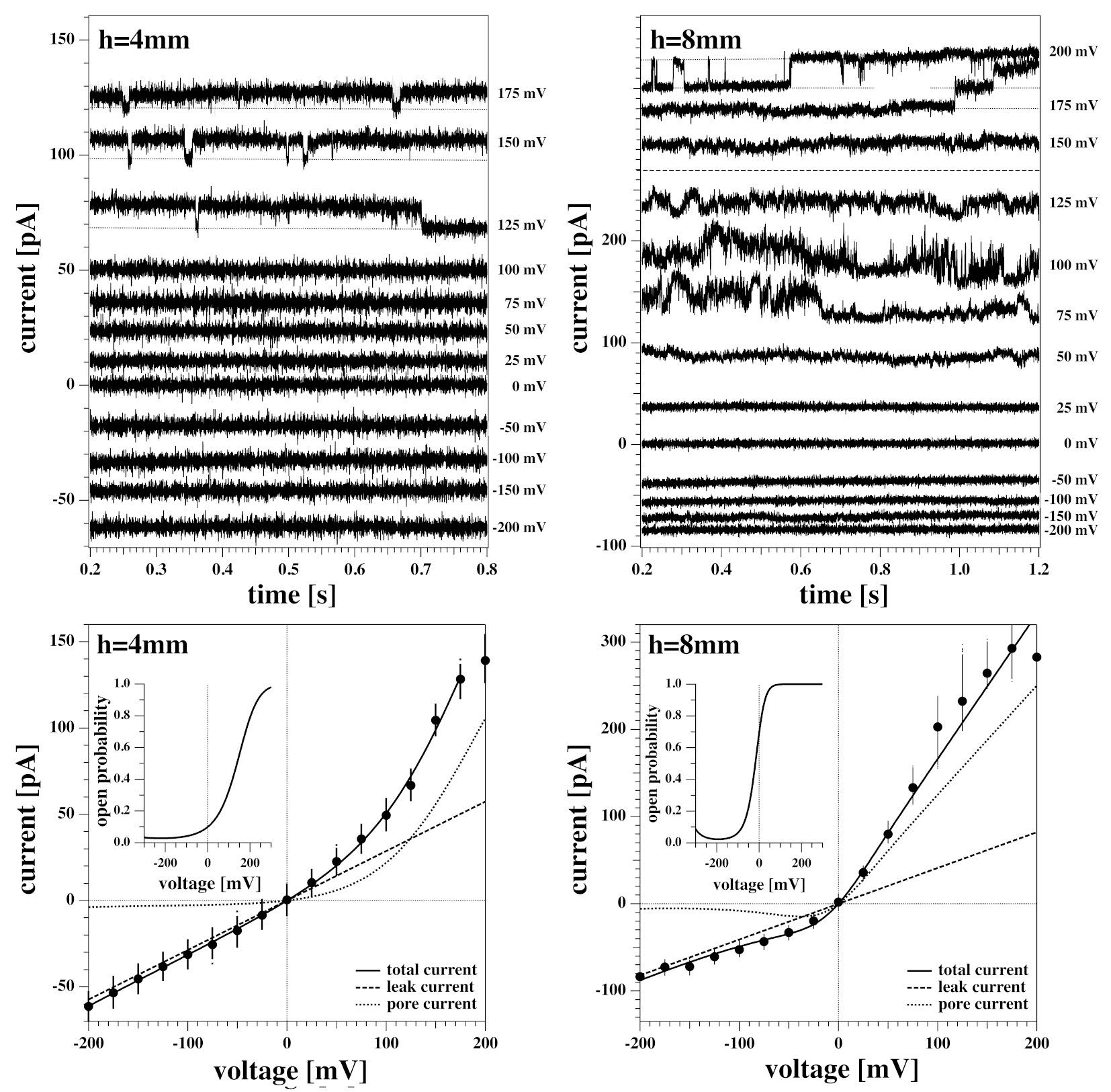}
	\caption{\small\textit {Current response of membrane 2 at $4\,$ mm (\textbf{Left}) and $8\,$ mm (\textbf{Right}) below the water surface. The current response to negative voltages doesn't show current fluctuations. \textbf{Left:} The membrane current at high positive voltages shows quantized steps to a lower current value before jumping down to it and staying there for the lower voltages. The current response at $200\,$ mV shows a constant drift in the baseline and is not shown here. \textbf{Right:} The current responses to $200\,$, $175$ and $150\,$ mV have been shifted to the top panel for clarity as they overlap. The membrane shows current fluctuations for increasing voltages. They end up in well-defined quantized steps at high voltages when pores are open most of the time.}}
	\label{figure7}
\end{figure*}

Fig.\ref{figure7} shows two current responses of membrane 2 to the same voltage protocol as describe in section \ref{electrophysiologicalexperiments}. The membrane had the same lipid composition as the one of Fig.\ref{figure4} - \ref{figure6}. The two recordings shown here were obtained with the membrane at $4\,$mm (left) and $8\,$mm (right) below the water surface.

One can recognize that the I-V profiles in both cases are non-linear. They display a larger conductance at positive voltages, which is more pronounced at the depth of 8 mm as compared to the 4 mm recording. Further, one recognizes discontinuous conduction steps in both experiments, which are also more pronounced at 8 mm depth. The fits to the I-V profile are performed as in Figs. \,\ref{figure5} and \,\ref{figure6}.

\begin{figure}[htbp]
	\centering
	\includegraphics[width=226pt,height=171pt]{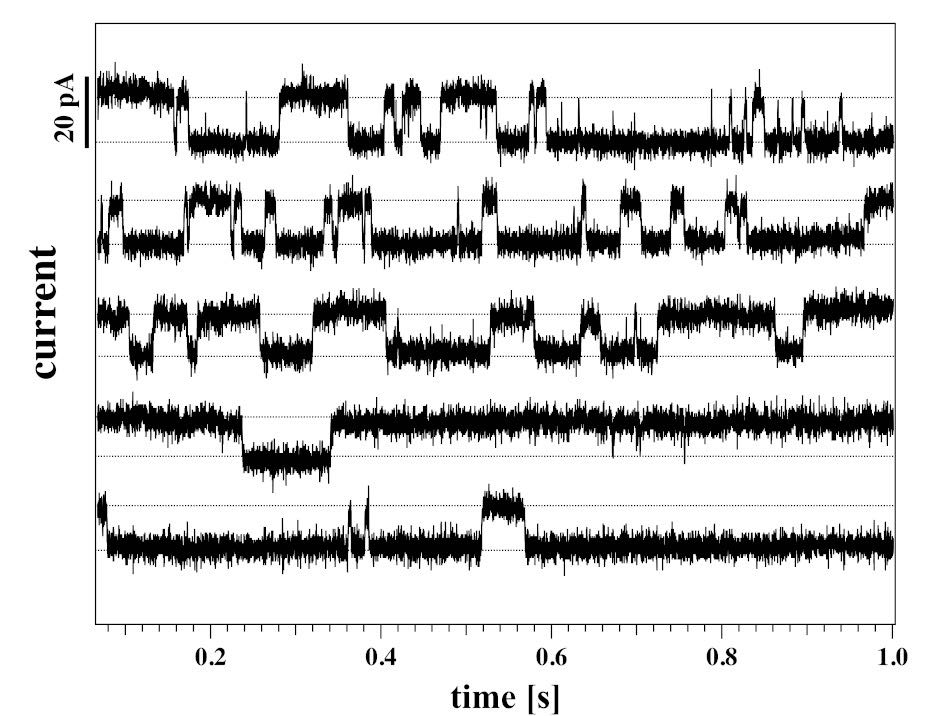}
	\caption{\small\textit {Continuous current recording for the same membrane as Fig. \ref{figure7} (membrane 2) at a holding voltage of $190\,$mV 4mm below the water surface ($T=297.15\,$K). The graph shows 5 consecutive seconds of recording, starting from the top row. The time-intervals are presented as a stack for clarity.}}
	\label{figure8}
\end{figure}

A continuous recording of this membrane at 4 mm depth and a fixed holding voltage allowed to observe consistent \linebreak channel-like activity for a longer period of time. Fig. \ref{figure8} shows the first $5\,$s of the raw current trace. The trace shows clear quantized steps from a baseline at $94\,$pA that was slightly drifting to a set value of about $106\,$pA. The step-size was about 13 pA, corresponding to a single lipid channel conductance of about $\sim 68$ pS, very similar to the single channel conductance in the same experiment at 150 mV and 125 mV. Thus, the single channel conductance is probably independent of voltage, as already found in \cite{Blicher2013}. The step size at 200 mV for the 8 mm depth recording in Fig. \,\ref{figure7} (right, top) yields a single channel conductance of about 137 pS, which is twice as high as in the 4 mm recording in the left panels. The current traces in Fig. \ref{figure7} indicate that at high positive voltage, pores can be open most of the time.

The above recordings were observed when the pipette tip was $4\,$mm and $8\,$mm below the water surface, so slightly lower than the recordings shown for membrane 1 in Fig.\ref{figure6}.

\begin{figure}[htbp]
	\centering
	\includegraphics[width=250pt]{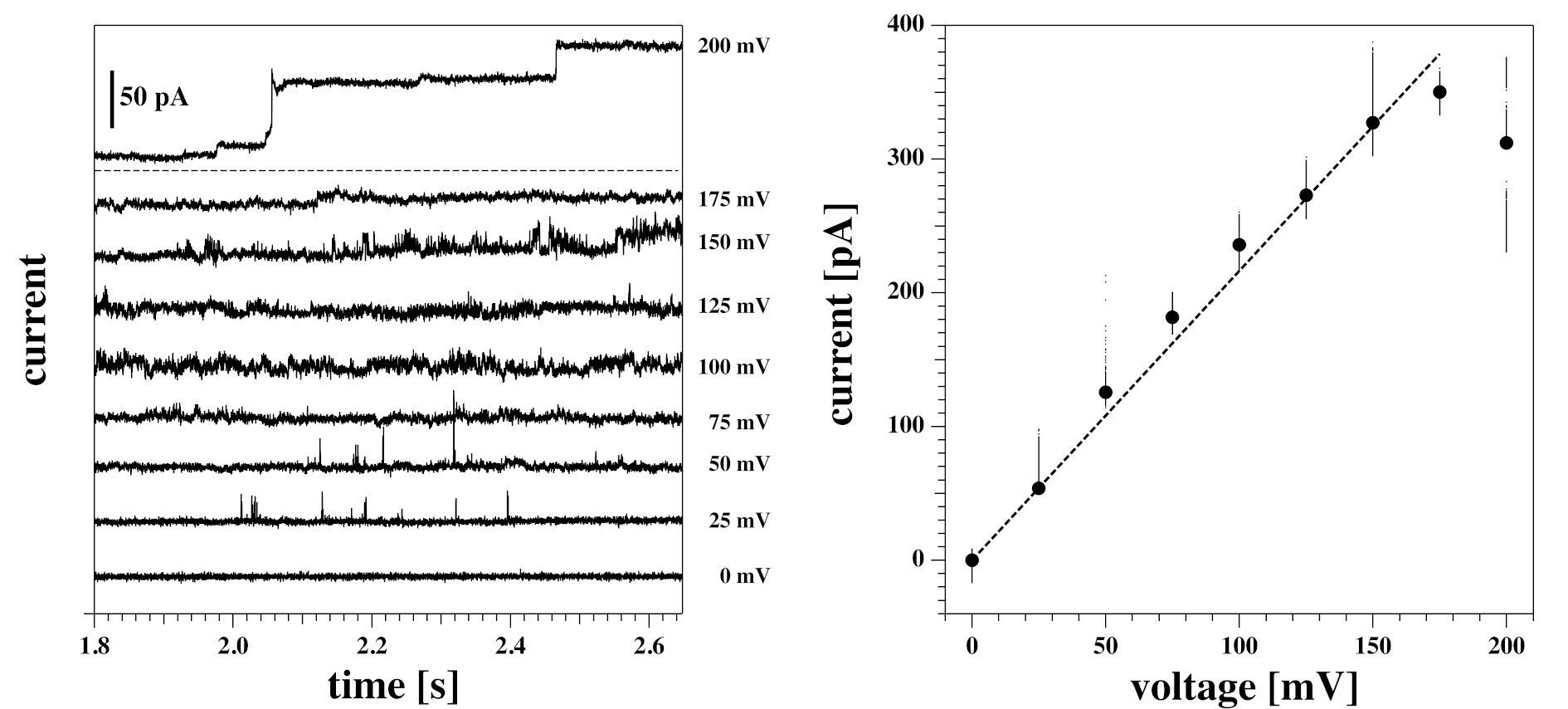}
	\caption{\small\textit {Current response of the same membrane as Fig.\ref{figure5}-\ref{figure6} (membrane 1) at a depth of 10 mm below the water surface. The membrane ruptured at $0\,$mV. Hence only the positive range is shown. \textbf{Left}: Current traces at different voltages. The top trace is the response to the first jump of the protocol (from $0$ to $200\,$mV), and it has been plotted separately for clarity. At this voltage, the membrane shows a step-wise increase in conductance, from an initial value of about $1.2\,$nS to a final one of roughly $1.8\,$nS. In the bottom panel, the other traces are shown. Current fluctuations start to appear in the form of spikes already at $50\,$mV, and increase in number and duration at higher voltages. \textbf{Right}: Current-voltage relationship for the traces shown at the left. A linear fit gives a value of the conductance of $1.8\,$nS. This is in line with the final value of conductance for the trace at $200\,$mV. Due to its transient behaviour, however, this trace has not been included in the fit. }}
	\label{figure9}
\end{figure}

We observed channels at high positive voltages and at different depths. In membrane 2 (Fig. \ref{figure7} and Fig. \ref{figure8}), channel activity seemed to increase with increasing depth of the tip in the water bath. At the same time, the voltage threshold for activity onset seemed to decrease with increasing depth\slash pressure. However, we have only three data points in this sense (i.e. three different depths). Note that the depths here are larger than those for the previous membrane. Interestingly, also the first membrane showed current fluctuations when brought $10\,$mm below the air-water interface. This is shown in Fig. \ref{figure9}. The current traces correspond to voltage steps in the range $200-0\,$mV (we did not record traces for negative voltage because at 0 V the membrane ruptured).

The membrane in Fig.\ref{figure9} ruptured at $0\,$ mV. Indeed, membrane rupture at large depths was one of the main obstacles to investigating this phenomenon further. In fact, membrane instability seemed to increase with increasing depth.

Summarizing, it seems that larger depth of the pipette enhances channel activity, and they occur at lower voltage. In the previous section, we have also found large conductances $g_L$ and $g_p$ at larger depth. This suggests that the lipid membrane channels are mechanosensitive.

\subsubsection{Symmetric I-V profiles in black lipid membranes}
\label{symmetrici-vprofilesinblacklipidmembranes}

In a previous publication \cite[]{Zecchi2017} we suggested that the offset voltage responsible for the asymmetry and the apparent outward rectification of the I-V profiles is caused by flexoelectricity. Bending of the membrane creates an electrical polarization that is roughly proportional to the curvature \cite[]{Petrov2001, Petrov2002a, Mosgaard2015b}. The tip diameter of a patch pipette is small. The maximum curvature possible is that of a half sphere with a radius that corresponds to the radius of the tip opening. For a pipette with 1 \textmu m diameter, the maximum curvature is $c=1/500 nm$. However, if one uses black lipid membranes spanning a hole with a diameter of about 100 \textmu m, the maximum possible curvature is 100 times smaller (see \cite[]{Zecchi2017} for details about this argument). If the voltage offset were in fact caused by flexoelectricity, the voltage offset would be practically absent in black lipid membrane measurements. This means that patch pipettes allow for high curvature, while black lipid membranes rather imply membranes without or with low curvature.

\begin{figure*}[htbp]
	\centering
	\includegraphics[width=393pt,height=185pt]{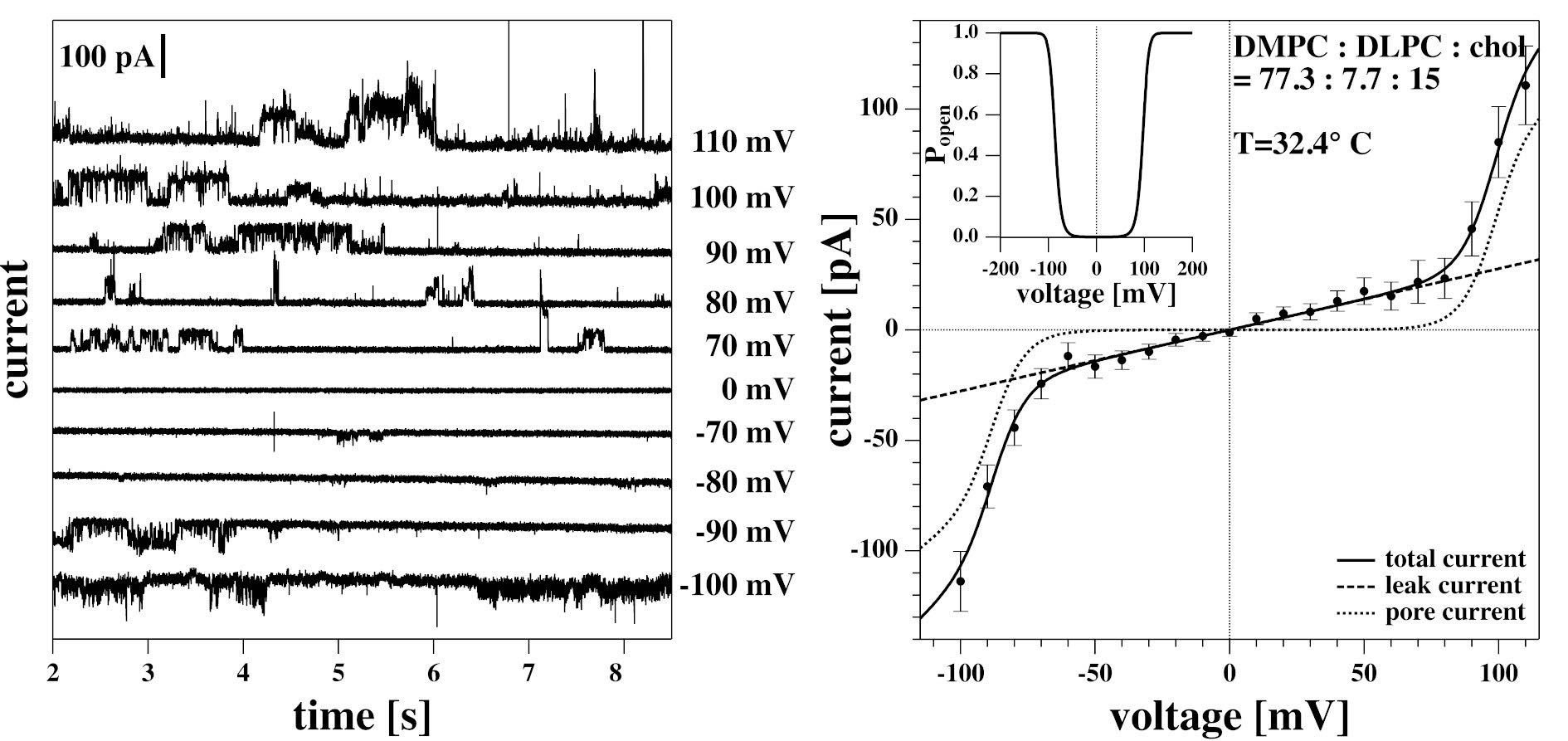}
	\caption{\small\textit {Recordings on a DMPC:DLPC:chol.= 77.3 : 7.7 : 15 BLM measured at 32.4$^\circ$ C. Left: Current traces. For better visibility, they are evenly spaced on the vertical axis. Note, the current traces in the linear regime of the I-V profile are not shown. Right: I-V-profile. The solid line is a fit of the I-V profile to eq. (\ref{eq:theor1.5}). The insert represents the pore open probability. Note that for the BLM, the I-V profile is symmetric and the channel activity occurs at both higher positive and higher negative voltages. }}
	\label{figure10}
\end{figure*}

Fig. \ref{figure10} shows a black lipid membrane experiment made with the Ionovation Explorer (see Material and Methods section). The aperture in this experiment had a diameter of about 120 \textmu m. In the patch experiment it was 8 $\sim$ \textmu m, i.e., about 15 times smaller. We used a membrane with 77.3 \% DMPC, 7.7 \% DLPC and 15 \% cholesterol and a temperature of 32$^\circ$ C at the upper end of the melting transition of this membrane (Fig. \ref{figure1}). Cholesterol is known for largely broadening the melting transitions of membranes, thus reducing domain formation and fluctuations. This renders membranes more stable. This is an important factor in black lipid membrane measurements because of the notorious instability of the membranes close to transitions.

We found an I-V profiles that is completely symmetric, i.e., it does not display a significant voltage offset (the fit yields $\sim$ -5 mV). We found a leak conductance of $g_L=270$ pS and a pore conductance of $g_p=863$ pS when fitting it with eq. (\ref{eq:theor1.4}). The conductance steps at 100 mV correspond to 78 pA, comparable to a single channel conductance of $\sim 78$ pS. This is similar to the single channel conductance of the traces in Fig. \ref{figure6} where we found $\sim 68$ pS. This indicates that the magnitude of the conductance steps does not depend on the size of the membrane. Most interestingly though, the near absence of a voltage offset allows seeing that the rectification pattern is also symmetric, and that channel activity appears both at positive and negative voltages in the nonlinear part of the I-V profile that we attribute to pore formation. This supports our interpretation of having two conduction processes present.

\subsubsection{Membrane relaxation}
\label{membranerelaxation}

\begin{figure*}[htbp]
	\centering
	\includegraphics[width=447pt,height=159pt]{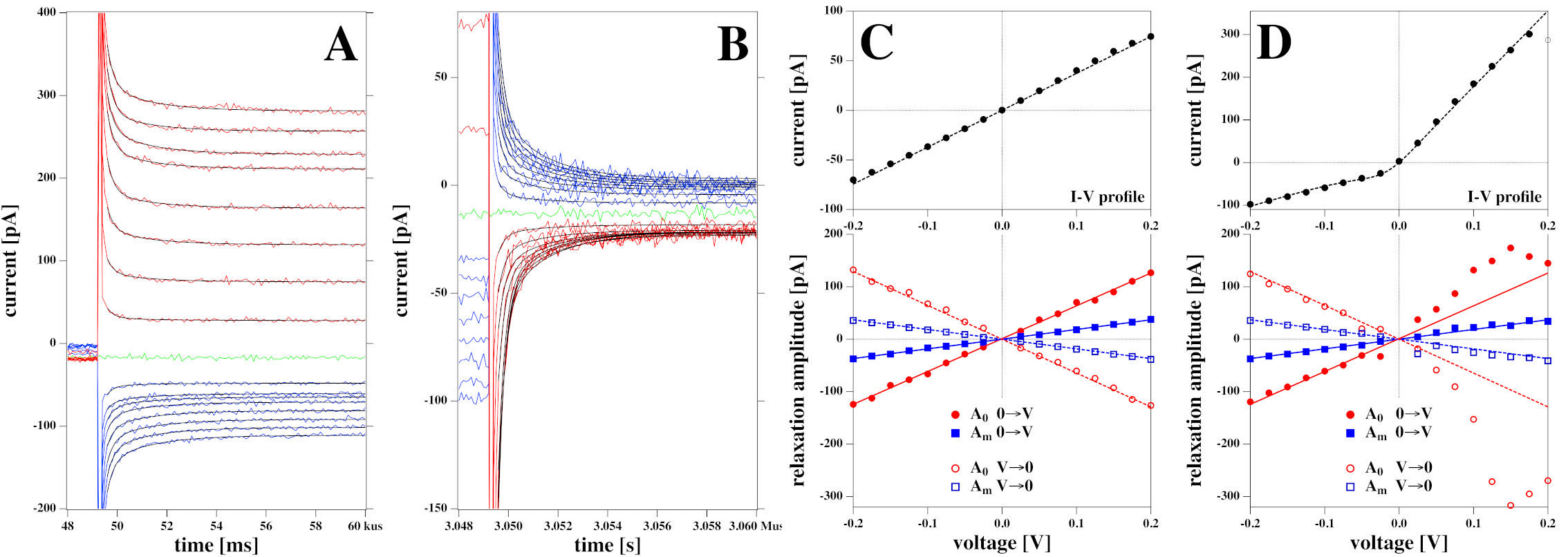}
	\caption{\small\textit {Relaxation processes corresponding to the I-V profiles shown in Fig. \,\ref{figure4}C and D: \textbf{A.} Relaxation processes after a jump from 0 volt to a fixed voltage for the rectified I-V profile shown in Fig. \,\ref{figure4} D. The thin black lines represent the fits to a biexponential function. V. \textbf{B.} Relaxation processes after jump from a fixed voltage V to zero volts for the rectified I-V profile shown in Fig. \,\ref{figure4} D. \textbf{C.} Bottom panel: The fit parameters of the relaxation processes corresponding to Fig. \,\ref{figure4}C. The top panel shows the corresponding I-V relation Both the amplitudes of the fast and the slow process, $A_0$ and $A_m$ are linear as is the IV profile. \textbf{D}. Bottom panel: The fit parameters of the relaxation processes corresponding to Fig. \,\ref{figure4}D. Here, $A_m$ is linear in voltage and equal in magnitude in both experiments (blue lines). Amplitude $A_0$ displays a nonlinear voltage dependence for the rectified I-V profile (right). The solid lines at the bottom are the same in the panel C and D. In the panel D they serve as a guide to the eye to demonstrate the nonlinearity of the voltage dependence for amplitude $A_0$. The current relaxation profiles in panels A and B were not corrected for a slight current offset from zero.}}
	\label{figure11}
\end{figure*}

In section \ref{relaxationprocesses} we described how after a voltage jump one finds transient equilibration processes that consist of capacitive currents and time-dependent changes in conductance (Fig. \,\ref{figure3}). We have argued that one expects (at least) two relaxation processes. One is related to charging the capacitor via the electrolyte and a second one is coupled to changes in membrane structure with effects both on conductance and capacitance. In the present set of experiments, we performed voltage jumps from zero to a fixed voltage, and back to zero. The voltage across the membrane must not necessarily be equal to the voltage adjusted by the setup. The resistance of the pipette is a series resistance of glass pipette and membrane. Occasionally, our membrane ruptured during the experiment (e.g., Fig. \,\ref{figure7}). Thus, one can observe the resistance of the setup with and without membrane after an instantaneous process. We found that in the absence of a membrane, the resistance is at least 2 orders of magnitude smaller than in the presence of the membrane. From this we concluded that the adjusted voltage was practically identical to the voltage across the membrane and that the resistance and capacitance of the pipette (with a tip diameter of about 8 \textmu m) do not contribute much to the observed currents.

We describe the relaxation process by a bi-exponential function (see Fig. \,\ref{figure3} and Theory section):

\begin{equation}
I(t) = g_a V_a+ A_0 e^{-\frac{t}{\tau_0}} + A_m e^{-\frac{t}{\tau_m}} \nonumber
\end{equation}

We fitted the relaxation profiles shown in Fig. \,\ref{figure11} A and B with biexponential profiles. We found that one obtains good fits of the relaxation profiles if the same relaxation times were taken for a given membrane for all voltages jumps from zero to a fixed voltage (Fig. \,\ref{figure11} A, 0$\rightarrow$V) and back (Fig. \,\ref{figure11} B, V$\rightarrow$0). Empirically, we adjusted the two relaxation times to 258 \textmu s (dominated by the digital filter time constant of the experiment) and 1.74 ms for all profiles.What is fitted are the amplitudes $A_0$ and $A_m$ of the two relaxation processes (Fig. \,\ref{figure11} C and D).

We find that the bi-exponential relaxation profiles with fixed time constants describe the experimental data well. The amplitude $A_m$ of the slow process displays a linear dependence on the voltage for both the linear and the rectified I-V-relation shown in the top panels of Fig. \,\ref{figure11} C and D (blue symbols). In contrast, the fast process with amplitude $A_0$ (red symbols) is similar to the voltage dependence of the IV relation. It displays a linear behavior for the linear IV relation in Fig. \,\ref{figure11} C, and a rectified behavior for the rectified I-V relation in Fig. \,\ref{figure11} D. The analysis of the data in Fig. \,\ref{figure7} yields a similar result. Taking into account the similarities between the voltage-dependence of the fast process and the steady state current, it seems plausible to assume that the fast relaxation processes contains elements from the relaxation of the conductance after a voltage jump. Since we attributed the non-linearity of the I-V profile to the opening of pores in the Theory section, we suggest that the fast process is related to the timescale of pore opening.

\section{Discussion}
\label{discussion}

In this paper, we described the conductance of DMPC-DLPC=\linebreak 10:1 membranes subject to voltage jumps of different magnitude and direction. The experiments had different aspects: 1. Studying I-V profiles at different depths of a patch pipette in the aqueous buffer. 2. Studying discrete channel opening- and closing events. 3. Comparing BLMs with lipid patches 4. Determination of the kinetics of membrane equilibration after a voltage jump.

For each of the two membranes that were stable enough for an extended experimental sequence in the patch clamp experiments that we describe here, we found two quantitatively different current-voltage profiles. We either found linear I-V relationships or outward rectified I-V profiles, even under the same conditions and for the same membrane. We could well describe these profiles with a model that allows for two electrical conduction processes: A voltage-independent leakage and a voltage-dependent pore-formation (i.e., the occurrence of lipid ion channels). This is consistent with our observation that membrane channels are usually found in the rectified profiles at positive voltages. The asymmetry of the rectified I-V profile is a consequence of an offset potential $V_0$, i.e., a polarization of the membrane. The leakage-current of the rectified profiles was within error identical to that of the linear I-V profiles found for the same membrane in the absence of pores. One therefore has to conclude that the pore-formation sometimes lacks the nucleation sites for pore formation (sometimes called a pre-pore in the literature \cite[]{Boeckmann2008}). Sometimes pores are present and sometimes they aren't. Details of this somewhat stochastic process remain to be explored. In contrast, the leak current is always present. In our theory, we assumed that the voltage-dependence of the free energy of the pores is quadratic in voltage, which has been proposed earlier \cite[]{Winterhalter1987, Blicher2013, Mosgaard2015a}. In \cite{Winterhalter1987} it was also proposed that voltage may stabilize the formation of pores with a given radius, i.e., pores of fixed conductance, which we have also found here. Interestingly, the I-V relations of the BLM measurement was nonlinear but symmetric, in agreement with previous findings \cite[]{Wodzinska2009}.

In the past, we have attributed an offset potential of the membrane to a spontaneous polarization of the membrane. Since our lipids are uncharged, the polarization may arise from membrane curvature due to an effect called flexoelectricity \cite[]{Petrov2006, Mosgaard2015a}. It arises from the asymmetric distribution of lipid dipoles in curved membranes. It may also arise from the asymmetric attachment of membranes to the glass pipette or from a lipid asymmetry (not likely in the present experiments). The offset potential $V_0$ results in the rectified profiles. The voltage dependence of the pore open probability is also relative to the voltage $V_0$.

We have changed the depth of the pipettes in the aqueous medium. This changes the pressure difference and might potentially lead to a change in the offset potential. For membrane 1 we found 243 ( $\pm$ 34) mV on average in 1 mm depth, while we found 221 ( $\pm$ 20) mV on average at a depth of 3 mm. The difference of these two values of $V_0$ is smaller than the standard deviation. Similarly, for membrane 2 we found $V_0=221$ mV at 4 mm depth and $V_0=193$ mV at 8 mm depth. Thus, while it might be that the offset potential is smaller at larger depth, the error margin does not allow us to make a trustworthy statement about its depth dependence. It seems that the origin of membrane polarization is not primarily the pressure difference across the membrane. However, it was generally true that the leak conductance $g_L$ and the pore conductance $g_p$ were larger at larger depth of the patch pipette. For membrane 1, the conductance $g_L$ increased by 32\% and the pore conductance $g_p$ by 37\%. For membrane 2, $g_L$ change by 43\% upon going from 4mm to 8mm depth, and $g_p$ changed by 86\%, respectively.

Interestingly, one does not find a measurable offset potential in black lipid membranes (Fig. \ref{figure7b}). We attribute that to the fact that the maximum possible curvature for a membrane across an aperture of 120 \textmu m in our BLMs is much smaller that that of a lipid patch spanning a tip aperture of about 8 \textmu m.

We reported that the rectified I-V relations are often accompanied by the voltage-gated opening of single lipid ion channels (see also \cite[]{Laub2012, Blicher2013, Mosgaard2013b}). This is surprising because it is usually believed that voltage-gated conduction-events are an exclusive feature of protein channels. In our experiments, the minimum pore open probability is found for $V-V_0=0$, i.e., at a voltage around -200 mV. One expects opening of pores below about -400 mV (outside of our experimental range) and above about 0 V. This has in fact been observed in Fig. \,\ref{figure7}. In the BLM measurements that do not display a voltage offset, the formation of channel events displays symmetric voltage-dependence, i.e., it occurs at both higher positive and negative voltages. In patch experiments, we found a single channel conductance at 8 mm depth that was 2.0 times larger than the single channel conductance at 4 mm depth ($\gamma=$ 137 pS versus $\gamma=$ 68 pS, respectively). In the BLM experiment, the single channel conductance was $\gamma=$ 78 pS.

In the previous section, we outlined that the conductance $g_p$ of the I-V profile associated with open pores was about 2 times larger at 8 mm depth than at 4 mm depth. These numbers are well in agreement with the difference in single channel conductance. This indicates that interpreting a conduction process related to pore-formation is reasonable. The second conduction process that we called a leak-current is voltage-independent and does not display discrete channel events. Larger depth of the pipette enhances channel activity, and they occur at lower voltage. This suggests that the lipid membrane channels are mechanosensitive.

The conductance of the lipid pores is of a magnitude typical for single channel proteins. \cite{Llano1988} reported single potassium channel conductances of $\gamma=$ 10, 20 and 40 pS in the squid axon. \cite{Salkoff2006} found that the so-called SLO-potassium channel family may have larger conductances, e.g. $\gamma=$ 100--270 pS (SLO1), $\gamma=$ 60--140 pS (SLO2.1), $\gamma=$ 100--180 pS (SLO2.2) and $\gamma=$ 70--100 pS (SLO3). \cite{Sakmann1984} described a potassium channel with a single channel conductance of 27 pS. Below we show that also the voltage-gating properties of our membranes are similar to potassium channels. This suggests that the phenomenology of lipid and protein channels is practically indistinguishable.

After a voltage jump one finds a transient time regime of about 10 ms, in which the membrane equilibrates to a new state. In the theory section we outlined that one expects at least two relaxation processes. One of them is the charging of the membrane (and pipette) capacitor with a timescale independent of the membrane state, and the second one is a structural variation of the membrane leading to changes in capacitance, polarization and membrane conductivity. The second timescale reflects relaxation processes in the membrane. We could describe all current traces with bi-exponential fits containing a fast timescale of about 300 \textmu s (dominated by the low pass-filter time constant of the experiment) and a slow timescale of about 2 ms. The amplitude of the slow process was linear with the magnitude of the voltage jump and did not depend on whether the I-V profile was rectified or linear. In contrast, the amplitude of the fast process reflected the voltage-dependence of the steady-state conductance of the membrane after the voltage jump. We conclude that there exists a fast process altering the conductance by changing the membrane state. This could be reflected in a timescale of channel opening or of the overall membrane area. Usually one would assume that the membrane-related process is the slow process, which is opposite to the experimental evidence showing that the fast process reflects that rectification. The reasons for this remain to be explained. It is further not clear why the I-V profiles are sometimes rectified and sometimes not.

\textbf{Comparison of membrane conductance with the potassium channel of Hodgkin and Huxley}

We have found that many of the I-V curves are rectified, which we we explained by the voltage-gated opening of pores. These pores can be recognized in many of the current traces. Our analysis allows for a determination of the probability of pore opening. We found that close to the offset-voltage $V_0$, the pores are mostly closed, while the open probability increase as a function of $(V^2+2V V_0)$.

Rectified behavior has usually been attributed to voltage-gated protein ion channels. These channels were originally introduced by Hodgkin and Huxley in order to explain the properties of the nervous impulse in squid axons \cite[]{Hodgkin1952b}. In their model they introduced two channels (or more accurately: two gating mechanisms) for the conduction of sodium and potassium. The conductances of the the potassium channel, $g_K$, and of the sodium channels, $g_{Na}$ were described by

\begin{eqnarray}
g_K&=&g_{K,0}\cdot n^4(V,t)\nonumber\\
g_{Na}&=&g_{Na,0}\cdot m^3(V,t)\cdot h(V,t)
\label{eq:HH1}
\end{eqnarray}

where $n(V,t)$, $m(V,t)$ and $h(V,t)$ are voltage and time-dependent functions describing single-exponential kinetics of gate-opening in the channels. The potassium channel is described by 4 independent and identical gates. After a voltage jump from $V_0$ to $V$, the conductance of the potassium channel is described by

\begin{eqnarray}
n(V,t) &=& n_{\infty}(V) -(n_{\infty}(V)\nonumber\\
&&-n_\infty(V_0))\cdot\exp{\left(-\frac{t}{\tau_n(V)}\right)}
\label{eq:HH2}
\end{eqnarray}

where the relaxation time $\tau_n(V)$ and the steady state values of $n$, $n_\infty(V)$ and $n_\infty(V_0)$ are given by

\begin{eqnarray}
n_\infty(V) &=& \frac{\alpha_n(V)}{\alpha_n(V)+\beta_n(V)}\quad\mbox{and} \nonumber\\
\tau_n(V)&=&\frac{1}{\alpha_n(V)+\beta_n(V)}
\label{eq:HH3}
\end{eqnarray}
with
\begin{eqnarray}
\alpha_n(V) &=& \frac{10^4\cdot(V+0.055)}{1-\exp\left(-\frac{V+0.055}{0.010} \right)}\quad\mbox{;}\nonumber\\
\beta_n(V)&=&125\cdot \exp\left(-\frac{V+0.065}{0.08}\right)\nonumber
\label{eq:HH3}
\end{eqnarray}
where the voltage $V$ is given in units of {[V]}.\footnote{Note that in \cite{Hodgkin1952b} the voltage was defined relative to the resting potential.} The functions $\alpha_n(V)$ and $\beta_n(V)$ were not derived from first principles but rather parametrized from experimental voltage-clamp data \cite[]{Hodgkin1952, Hodgkin1952a}. The steady state open probability of the potassium channel is then given by

\begin{equation}
P_{open, K} (V)= n_\infty^4(V)
\label{eq:HH4}
\end{equation}

\begin{figure}[htbp]
	\centering
	\includegraphics[width=250pt]{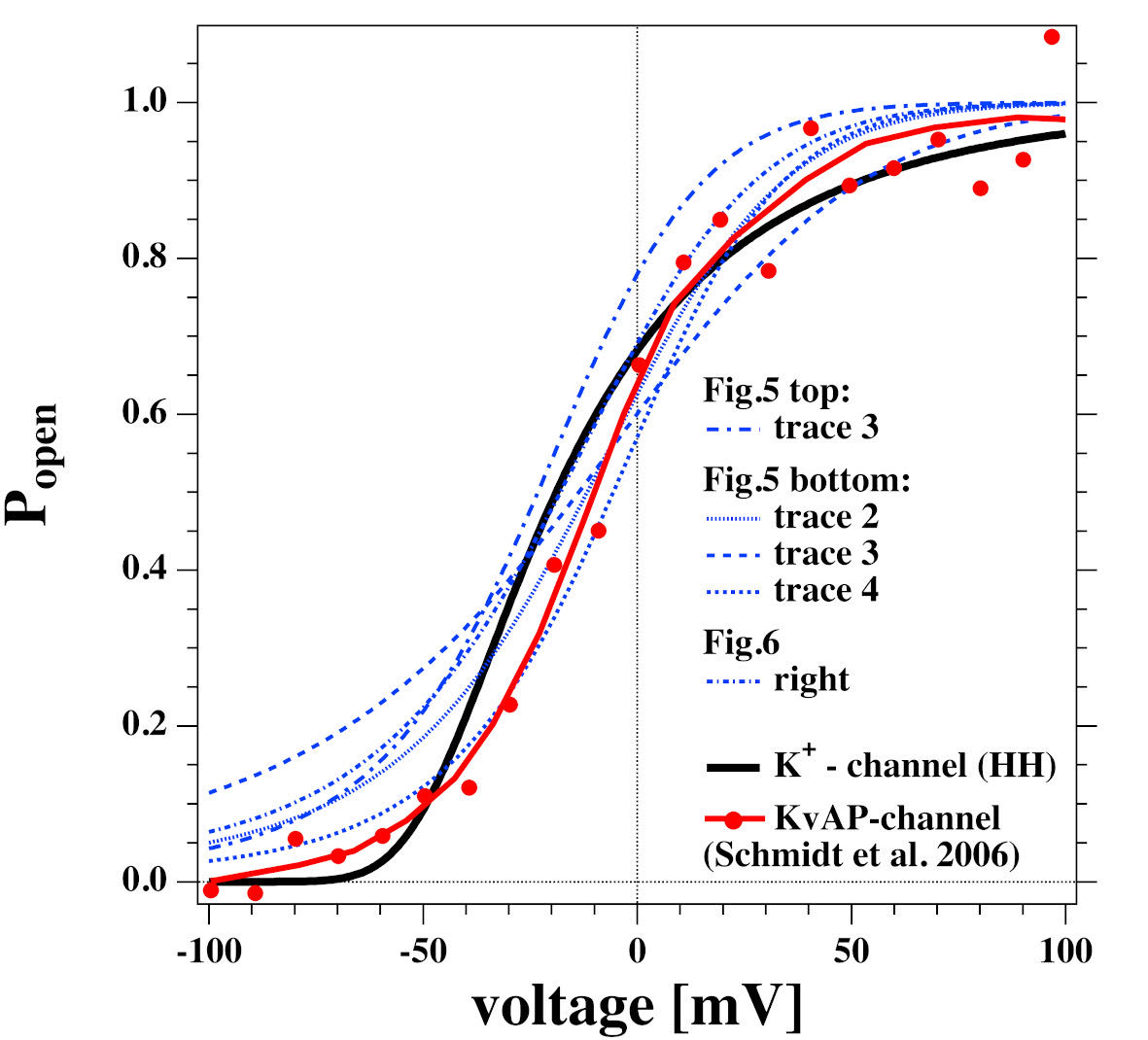}
	\caption{\small\textit {Comparison of the open probability of the K$^+$-channel of \cite{Hodgkin1952b} (fat line), the KvAP channel reconstituted into a POPE\slash POPC=3:1 membrane \cite[]{Schmidt2006}, and selected open probabilities from Fig. \,\ref{figure6} and \,\ref{figure7} in the voltage regime of physiological relevance.}}
	\label{figure12}
\end{figure}

This function is shown in Fig. \,\ref{figure12} (solid black line). It is interesting to compare it to the open probability of the pure lipid membranes of the experiments described above (blue traces).

We took the calculated open probabilities of 5 selected traces from Figs. \,\ref{figure6} and \,\ref{figure7} and plotted them in comparison to the steady state open probability of the K$^+$-channel determined by Hodgkin and Huxley. While the curves are not identical, they are in fact quite similar. Fig.\,\ref{figure7}er shows that the open probability in fact goes along with the opening of pores. For comparison, we also added the normalized conductance of the voltage-gated KvAP channel (red symbols, \cite{Lee2005, Schmidt2006}) that displays a voltage-dependent conductance that is very similar to the HH-potassium channel and the lipid channels from this publication. This demonstrates that many properties of the potassium channel can be described be the rectified properties of the lipid membrane in the absence of any macromolecules. The similarity of lipid membrane behavior with that of potassium channels has in fact been noticed before \cite[]{Blicher2013, Mosgaard2013b}.

Interestingly, \cite{Seeger2010a} showed that the both mean conductance and open channel life-times of the KcsA potassium channels reconstituted into a synthetic lipid membrane are proportional to the heat capacity of the lipid membrane. This behavior would be a logical consequence of membrane fluctuations if the membrane conductance originated from the lipid membrane alone. Since the conductance of the membrane was still proportional to the concentration of K-channels in the membrane, we have proposed in the past that the conductance in the membrane originates from membrane pores that are catalyzed by the channel protein \cite[]{Mosgaard2013b}.

\section{Conclusion}
\label{conclusion}

We describe the voltage-gated opening of channels in pure lipid membrane. When the I-V profiles are outward rectified, in most cases one finds single channel events at high voltage. When I-V profiles are linear, one does not see single channels. I-V profiles in BLMs are symmetric and show lipid channels at higher voltage. We concluded that there are two conduction processes: One leak current not related to membrane channels and with a conductance that is independent of voltage, and single channel events that are voltage-gated. We reported evidence for that channel-opening is mechanosensitive. The properties of the membrane channels resemble those of potassium channels.

\small{

}

\end{document}